# Visualization of Anomalous Ettingshausen Effect in a Ferromagnetic Film: Direct Evidence of Different Symmetry from Spin Peltier Effect

T. Seki,[1,2,*] R. Iguchi,[3] K. Takanashi,[1,2] and K. Uchida[2,3,4,†]

[1]*Institute for Materials Research, Tohoku University, Sendai 980-8577, Japan*

[2]*Center for Spintronics Research Network, Tohoku University, Sendai 980-8577, Japan*

[3]*National Institute for Materials Science, Tsukuba 305-0047, Japan*

[4]*PRESTO, Japan Science and Technology Agency, Saitama 322-0012, Japan*

**Abstract:**

Spatial distribution of temperature modulation due to anomalous Ettingshausen effect (AEE) is visualized in a ferromagnetic FePt thin film with in-plane and out-of-plane magnetizations using the lock-in thermography technique. Comparing the AEE of FePt with the spin Peltier effect (SPE) of a Pt / yttrium iron garnet junction provides direct evidence of different symmetries of AEE and SPE. Our experiments and numerical calculations reveal that the distribution of heat sources induced by AEE strongly depends on the direction of magnetization, leading to the remarkable different temperature profiles in the FePt thin film between the in-plane and perpendicularly magnetized configurations.

(99 words)

\* e-mail: go-sai@imr.tohoku.ac.jp, † e-mail: UCHIDA.Kenichi@nims.go.jp

Spin-current ($J_s$)-mediated interconversion between electric charge current ($J_c$) and heat current ($J_q$), research field of which is frequently called "spin caloritronics" [1], has largely fascinated us from the viewpoint of not only fundamental physics but also potential applications. Spin Seebeck effect (SSE) [2-4] and spin Peltier effect (SPE) [5,6] are representative phenomena of spin caloritronics. SSE enables us to convert a temperature gradient ($\nabla T$) to pure $J_s$ owing to the collective magnetization dynamics activated by $\nabla T$ [7,8]. On the other hand, SPE is the reverse process of SSE, in which the flow of $J_s$ produces $J_q$ due to the transfer of spin angular momentum and energy from conduction electron spins to local spins, and vice versa, and the resultant non-equilibrium states of the magnon and electron systems [5]. Eventually, the SPE induces $\nabla T$ along $J_s$. In both cases of SSE and SPE, junctions consisting of ferromagnetic and paramagnetic materials are often studied [2,3,5,6,9], *e.g.* a ferrimagnetic insulator yttrium iron garnet (YIG) and a paramagnetic metal Pt. For SSE, the $J_s$ due to the non-equilibrium spin state at the ferromagnet / paramagnet interface is observed as electric voltage in the paramagnet via the spin-orbit interaction, *i.e.* a spin Hall effect (SHE) [10]. The SPE has recently been observed in junctions with YIG and Pt by using microfabricated thermopiles [5] and active infrared emission microscopy called lock-in thermography (LIT) [6,11,12]. $J_s$ was generated via the SHE of Pt by the $J_c$ flow, and the interaction of $J_s$ and spontaneous magnetization ($M$) of YIG modulated the temperature of the junction.

In addition to the SSE and SPE, anomalous Nernst effect (ANE) and anomalous Ettingshausen effect (AEE) are famous thermoelectric phenomena in ferromagnets that have been known for a long time [13], in which the coexistence of $M$ and $\nabla T$ produces anomalous Nernst voltage while that of $J_c$ and $M$ leads to $\nabla T$ due to the AEE. The ANE has widely been studied for a variety of ferromagnetic bulks and thin films [14-23]. In contrast to ANE, there is no report on the observation of the AEE in a ferromagnetic thin film although a few studies on the AEE in ferromagnetic bulks have been reported [24-26]. The details of AEE such as magnetic field dependence have not totally been understood even for the bulks. The observation of AEE in a ferromagnetic thin film is particularly crucial in order to discuss the difference from the other spin caloritronic phenomena often observed in thin film devices. For example, one may be aware of the similarity of AEE and SPE, where $J_c$ leads to $\nabla T$ under $M$ for both cases. However, both effects should have different symmetries with respective to the relationship between $J_c$ and $M$ because the spin-polarization vector $\sigma$ of $J_s$ is another important quantity for the SPE [6]. Thus, the observation of AEE is a key to obtain definite experimental evidence of the different symmetries AEE and SPE should have.

We report the observation of AEE in a ferromagnetic FePt thin film by using the LIT. The temperature profile induced by AEE is clearly changed by varying $M$ from the in-plane direction to the out-of-plane direction, which are called "in-plane magnetized (IM)" and "perpendicularly magnetized (PM)" configurations. Different thermal images for AEE and SPE give us direct evidence of different symmetries for these phenomena. By comparing the experimental results with the numerical results calculated by the finite element method, also we semi-quantitatively discuss the magnitude of temperature change and the distribution of the heat source due to the AEE for the IM and PM configurations.

Two kinds of thin films are used in this experiment: a ferromagnetic FePt on a $SrTiO_3$ substrate for AEE and a paramagnetic Pt on a ferrimagnetic YIG for SPE. The 10-nm-thick FePt (100) layer was epitaxially grown on the $SrTiO_3$ (100) substrate at 350ºC using an ultrahigh vacuum magnetron sputtering system (See Supplementary Materials [27] for the details of structural characterization). Although the deposition at 350ºC led to the partially $L1_0$-ordered FePt and induced the non-negligible perpendicular magnetic anisotropy giving a small saturation field in the PM configuration, the easy magnetization axis of FePt is still in the film plane as shown later. On the other hand, a single-crystalline YIG (111) insulative layer with the thickness of 59 μm was prepared on a $Gd_3Ga_5O_{12}$ (111) substrate by a liquid phase epitaxy method. Then, the 10-nm-thick Pt layer was sputter-deposited onto the YIG at room temperature. The FePt and Pt films were patterned into the U-shapes (**Figs. 1(a) and 1(b)**) with the width of 500 μm using photolithography and Ar ion milling. **Figure 1(c)** is an optical microscope image of the U-shaped wire.

For thermal imaging of AEE and SPE, the LIT technique [6,11,12] was used. The infrared radiation thermally emitted from the sample surface was detected while applying rectangular-wave-modulated $J_c$ to the U-shaped FePt or Pt (**Fig. 1(d)**). The only first harmonic response was fed in order to separate the AEE or SPE contribution from the Joule heating contribution (**Fig. 1(e)**). To enhance infrared emissivity and to ensure uniform emission properties, the sample surfaces were coated with insulating black ink (emissivity > 0.95).

First let us consider the symmetries of AEE and SPE. In the case of AEE,

$$\boldsymbol{J}_\text{q}^\text{AEE} \propto \boldsymbol{J}_\text{c} \times \boldsymbol{M}^\text{FePt} \qquad (1)$$

is satisfied for both the IM and PM configurations, where the external magnetic fields are applied along the in-plane and the out-of-plane directions of the film ($H^\text{IP}$ and $H^\text{OP}$), respectively. As illustrated in **Fig. 1(a)**, $\boldsymbol{J}_\text{q}$ flows in the out-of-plane (in-plane) direction under $H^\text{IP}$ ($H^\text{OP}$) due to the symmetry of Eq. (1). For the SPE, the $\boldsymbol{J}_\text{s}$ in the Pt plays an important role and $\boldsymbol{\sigma}$ of $\boldsymbol{J}_\text{s}$ is directed along $\boldsymbol{J}_\text{c} \times \boldsymbol{n}$, where $\boldsymbol{n}$ is the normal vector of the interface plane. This $\boldsymbol{J}_\text{s}$ in the Pt interacts with $\boldsymbol{M}$ of YIG. Thus, the SPE satisfies [6,11,12]

$$\boldsymbol{J}_\text{q}^\text{SPE} \propto \left(\boldsymbol{\sigma} \cdot \boldsymbol{M}^\text{YIG}\right)\boldsymbol{n}. \qquad (2)$$

Since $\boldsymbol{\sigma}$ always lies in the film plane, we obtain the following relationship:

$$\boldsymbol{J}_q^\text{SPE} \begin{cases} \propto \boldsymbol{J}_\text{c} \times \boldsymbol{M}^\text{YIG} & \text{for IM configuration} \\ = 0 & \text{for PM configuration} \end{cases}. \qquad (3)$$

Namely, the IM configuration under $H^\text{IP}$ corresponds to $\boldsymbol{\sigma} \parallel \boldsymbol{M}^\text{YIG}$ whereas the application of $H^\text{OP}$ leads to $\boldsymbol{\sigma} \perp \boldsymbol{M}^\text{YIG}$. Therefore, we can experimentally examine the different symmetries of AEE and SPE by comparing temperature distributions between the IM and PM configurations.

The in-plane *M-H* curves for the FePt and the Pt / YIG under $H^\text{IP}$ are shown in **Figs. 2(a) and 2(b)**, respectively. Both samples have in-plane easy axes since high remanent magnetizations are obtained. **Figures**

**2(c) and 2(d)** display the amplitude ($A$) and phase ($\phi$) images at $J_c$ = 10 mA and the lock-in frequency ($f$) of 25.0 Hz for the FePt and Pt / YIG, respectively, where the amplitude and phase are defined in the range of $A \geq 0$ and $0° \leq \phi \leq 360°$. These images were obtained at $H^{IP}$ = 3.16 kOe, 0.06 kOe, and -3.14 kOe denoted by the solid circles in **Figs. 2(a) and 2(b)**. In the IM configuration, both the FePt and Pt / YIG exhibit the similar thermal images having the following characteristics: i) At $H^{IP}$ = 3.16 kOe, the temperature modulation, *i.e.* the increase of $A$ appears in the $y$-directional wires with the opposite $\phi$ at the right and the left wires, and ii) $\phi$ is reversed when $H^{IP}$ is reversed to -3.14 kOe. Namely, the sign of the temperature modulation is reversed by reversing $J_c$ or $H^{IP}$. This fact coincides with the symmetries of **Eq. (1)** and **Eq. (3)**. On the other hand, iii) no signal is detected in the region of $x$-directional wire. This is because $\boldsymbol{J}_c \parallel \boldsymbol{M}^{FePt}$ for FePt and $\boldsymbol{\sigma} \perp \boldsymbol{M}^{YIG}$ for Pt / YIG in the $x$-directional wires do not match the conditions of AEE and SPE, respectively. We confirm that the ordinary Ettingshausen effect in FePt and Pt is negligibly small by plotting the magnetic field dependence of $A$ and $\phi$ (See Supplementary Materials [27]). In the case of IM configuration, there is no remarkable difference in the thermal images between AEE and SPE as anticipated.

The significant difference can be seen in the thermal images of AEE and SPE for the PM configuration. The out-of-plane $M$-$H$ curves (**Figs. 2(e) and 2(f)**) indicate that the out-of-plane directions are the hard magnetization directions for both samples. The thermal images of FePt at $H^{OP}$ (**Fig. 2(g)**) show that the clear $A$ signals appear around the edges of U-shaped wire and the $\phi$ difference between the inside and the outside edges is ~ 180º. When $H^{OP}$ is swept from 7.84 kOe to -7.85 kOe, the reversed $\phi$ is observed. This temperature modulation over the entire edge is caused by AEE since $\boldsymbol{J}_c \perp \boldsymbol{M}^{FePt}$ is satisfied everywhere in the PM configuration. On the other hand, no signal is detected for the Pt / YIG with the PM configuration (**Fig. 2(h)** and also see Supplementary Materials [27]). This difference is direct evidence that AEE and SPE have the totally different symmetries (see **Eqs. (1) and (2)**). Also, we emphasis the following points: The YIG does not exhibit the AEE because of lack of conduction electrons in YIG, and we successfully get rid of the possibility of SPE signals contaminated by proximity-induced AEE in the Pt attached to YIG, which is consistent with the previous results on the SSE [28-30].

**Figures 2(i) and 2(j)** plot the $A$ signals on the samples as a function of $J_c$ for the IM and PM configurations, respectively. The $A$ signals for the IM configuration were obtained by averaging the data for the area enclosed by the white solid line shown in **Figs. 2(c) and 2(d)** while those for the PM were the maximum values in the averaged $x$-directional profile, which will be explained later in **Fig. 3**. For all the cases except the Pt / YIG at $H^{OP}$, the $A$ signal linearly responds to $J_c$, which is in agreement with the characteristics of AEE and SPE. For the IM configuration, the linear fitting to the data gives $A / j_c$, where $j_c$ is the charge current density, of $2.0 \times 10^{-13}$ Km$^2$/A and $3.1 \times 10^{-13}$ Km$^2$/A for the FePt and the Pt / YIG, respectively. $A / j_c$ for the Pt / YIG is comparable to a previous value [12]. It is noted that $A / j_c$ due to AEE for the FePt at $H^{IP}$ is of the same order as the SPE for the Pt / YIG at $H^{IP}$.

**Figure 3** shows the *x*-directional profiles of *A* and *ϕ* for the AEE of FePt at the IM ($H^{IP}$ = 3.16 kOe) and the PM ($H^{OP}$ = 7.84 kOe) configurations, respectively, where *f* was set at 25.0 Hz, 12.5 Hz, and 5.0 Hz. These *f*-varied images are useful to reveal the effect of thermal diffusion on temperature modulation. The temperature profiles of AEE for the IM configuration show abrupt transition between the region with and without the FePt wire regardless of *f*. These *f*-independent sharp temperature profiles indicate the steady-state condition of temperature modulation can immediately be achieved for the AEE at $H^{IP}$, which is similar to that of SPE (See Supplementary Materials [27]). On the other hand, the temperature profiles for the AEE with the PM configuration are different from the IM configuration. Although one anticipates the linear spatial distribution of temperature modulation along $J_q$ in a steady-state condition from **Eq. (1)**, the magnitude of temperature modulation nonlinearly decreases as the distance from the edges of FePt wire increases. In addition, the temperature profiles show the *f* dependence.

Here let us semi-quantitatively discuss the temperature change (Δ*T*) induced by AEE in FePt for the IM and PM configurations. Since $\Delta T \propto \int j_q \, dl$, where $j_q$ is the heat current density and *l* is the length, $\Delta T^{IM}$ for the IM configuration is proportional to the FePt thickness (*t* = 10 nm) while $\Delta T^{PM}$ for the PM configuration is proportional to the FePt wire width (*w* = 500 μm). Thus, $\Delta T^{PM} / \Delta T^{IM} = w / t = 5 \times 10^4$ is expected in ideal and isotropic system. In the present experiment, however, $\Delta T^{PM} / \Delta T^{IM} \sim 3.8$ is obtained at *f* = 25 Hz. In order to understand this significant discrepancy, we model the distribution of the heat source for AEE in FePt wires contacting the substrate and the black ink layer using the finite element method (See Supplementary Materials [27] for the details of modeling). **Figures 3(i) and 3(j)** show the calculated profiles of $A$ and $\phi$ for the IM and the PM configurations, respectively, where the values at the sample surface are shown and the calculation was performed with $f$ = 25.0 Hz. The calculated profiles of $A$ reproduce the experimental profiles for both the IM and the PM configurations. Also, the calculated frequency dependence of $A$ is qualitatively in agreement with the experimental results, and our modeling indicates that the large decrease in $\Delta T^{PM}$ can be reproduced by taking into account heat loss flowing from the FePt to the substrate and the black ink (See Supplementary Materials [27]). Importantly, the calculated $\Delta T^{PM} / \Delta T^{IM}$ is obtained to be ~ 4 at $f$ = 25.0 Hz where the thermal conductivity of FePt ($k^{FePt}$) is assumed to be 17.5 W m$^{-1}$K$^{-1}$. This $k^{FePt}$ value is of the same order as $k^{FePt}$ reported for a FePt film [31]. This means that the experimental $\Delta T^{PM} / \Delta T^{IM}$ can be reproduced by the calculation in the absence of the anisotropy of the AEE coefficient. In order to discuss the anisotropy of AEE, further investigation such as the accurate estimation of $k^{FePt}$ for the FePt thin layer is needed. Here, we emphasize the following thing. $\Delta T^{PM} / \Delta T^{IM}$ depends on the thermal property of the substrate. Since the factor for the Pt / YIG system is nearly the same as that for the FePt / SrTiO$_3$ considering the similar thermal properties of YIG to those of SrTiO$_3$ [32], the absence of $\Delta T^{PM}$ in the Pt / YIG is indeed due to the symmetry of the SPE.

In conclusion, the AEE of the FePt thin film was successfully observed for both the IM and PM configurations by exploiting the LIT technique. We compared the visualized temperature modulations between the AEE of FePt and the SPE of Pt /YIG, and showed the direct evidence for the different symmetries of AEE and SPE. It was found that the different distribution of the heat sources for the IM and PM configurations led to the different temperature profiles of AEE in the case of thin film structure. Our findings are crucial to develop thin film devices based on the spin caloritronic phenomena.

**Acknowledgement**

The authors thank S. Daimon for valuable discussions, and W. Zhou, Y. Murakami, and I. Narita for their technical support. This work was partially supported by Grant-in-Aid for Scientific Research (S) (JP25220910), Grant-in-Aid for Scientific Research (A) (JP15H02012) and Grant-in-Aid for Scientific Research on Innovative Area "Nano Spin Conversion Science" (JP26103005) from JSPS KAKENHI, Japan, PRESTO "Phase Interfaces for Highly Efficient Energy Utilization" (JPMJPR12C1) from JST, Japan, Collaborative Research Center on Energy Materials in IMR (E-IMR), and NEC cooperation. The structural characterization and the device fabrication were partly carried out at the Cooperative Research and Development Center for Advanced Materials, IMR, Tohoku University.

**Figure Captions**

**Figure 1** Illustrations of device structures for (a) anomalous Ettingshausen effect (AEE) consisting of a U-shaped FePt wire on a SrTiO3 substrate and (b) spin Peltier effect (SPE) consisting of a U-shaped Pt wire on a yttrium iron garnet (YIG). The top panels are bird's views while the middle and bottom panels are cross-sectional views under the in-plane magnetic field ($H^{IP}$) and the out-of-plane magnetic field ($H^{OP}$), respectively. $J_c$ is the a.c. charge current applied to the FePt or the Pt wire. $J_q$ and $J_s$ are the heat current and the spin current, respectively. (c) Optical microscope images of the fabricated device with the wire width of 500 μm. (d) Input and (e) output signals of lock-in thermography. A rectangular-waveformed a.c. charge current with amplitude of $J_c$ and frequency of $f$ was applied, and the first harmonic response was detected.

**Figure 2** Spatial distributions of temperature modulation for the in-plane magnetized (IM) and the perpendicularly magnetized (PM) configuration. (a) In-plane *M-H* curves for the FePt and (b) the Pt / YIG. (c) Amplitude ($A$) and phase ($\phi$) images for the IM configuration at $J_c$ = 10 mA and $f$ = 25.0 Hz for the FePt and (d) the Pt / YIG. $H^{IP}$ was set at 3.16 kOe, 0.06 kOe, and -3.14 kOe which are denoted by the solid circles in (a) and (b). (e) Out-of-plane *M-H* curves for the FePt and (f) the Pt / YIG. (g) $A$ and $\phi$ images for the PM configuration at $J_c$ = 10 mA and $f$ = 25.0 Hz for the FePt and (h) the Pt / YIG. $H^{OP}$ was set at 7.84 kOe and -7.85 kOe. (i) The $A$ signals on the samples as a function of $J_c$ for the IM configuration and (j) the PM configuration for the FePt and the Pt / YIG. The $A$ signals for the IM configuration were obtained by averaging the data for the areas enclosed by the white solid lines shown in (c) and (d) while those for the PM were the maximum values in the averaged $x$-directional profile.

**Figure 3** $x$-directional profiles of $A$ and $\phi$ for the AEE of FePt for the IM and the PM configurations, where the values were obtained by averaging the data for the hatched areas in (a) and (e). $f$ was set at (b) 25.0 Hz, (c) 12.5 Hz, and (d) 5.0 Hz for the IM configuration and at (f) 25.0 Hz, (g) 12.5 Hz, and (h) 5.0 Hz for the PM configuration. The external magnetic field was set at $H^{IP}$ = 3.16 kOe and $H^{OP}$ = 7.84 kOe for the IM and PM configurations, respectively. The calculated profiles of $A$ and $\phi$ for (i) the IM and (j) the PM configurations using the finite element method, where the values were obtained at the surface of black ink and the calculation was performed with $f$ = 25.0 Hz.

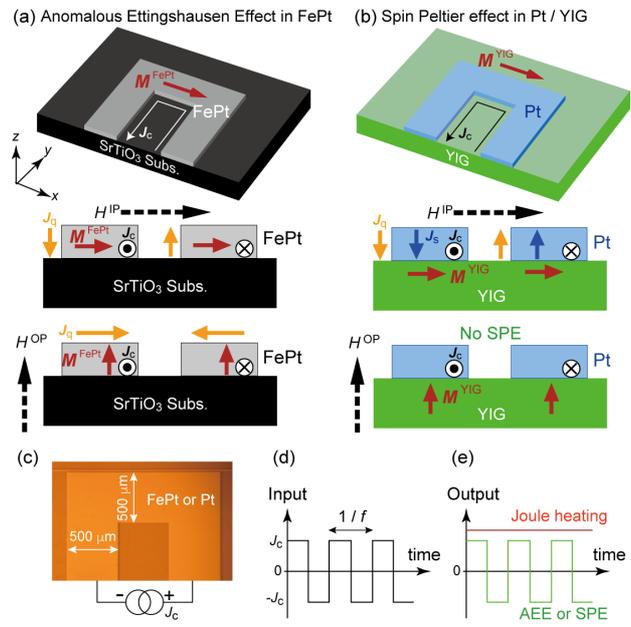

Figure 1

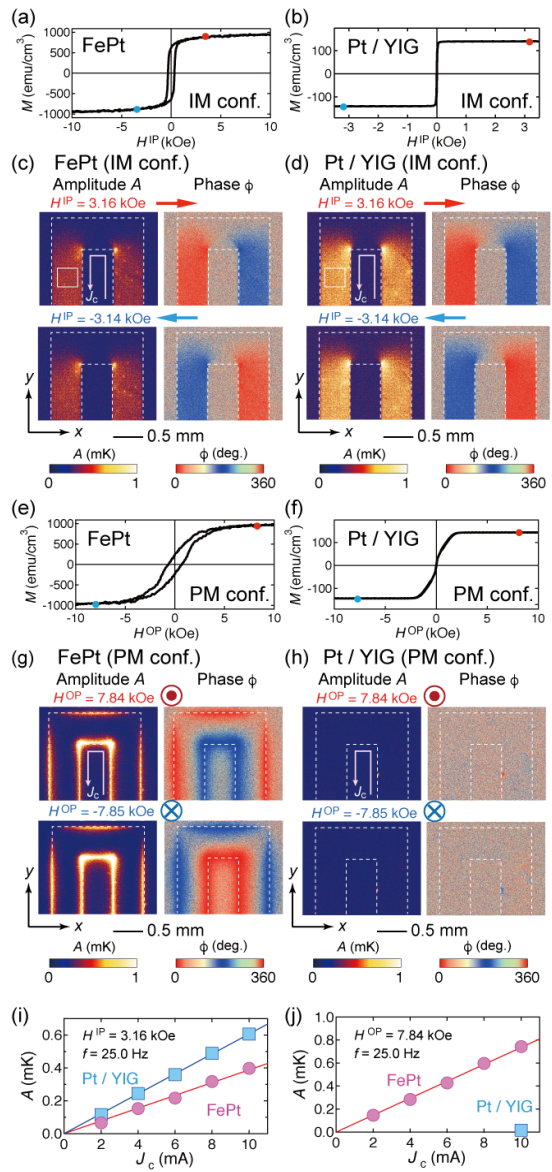

Figure 2

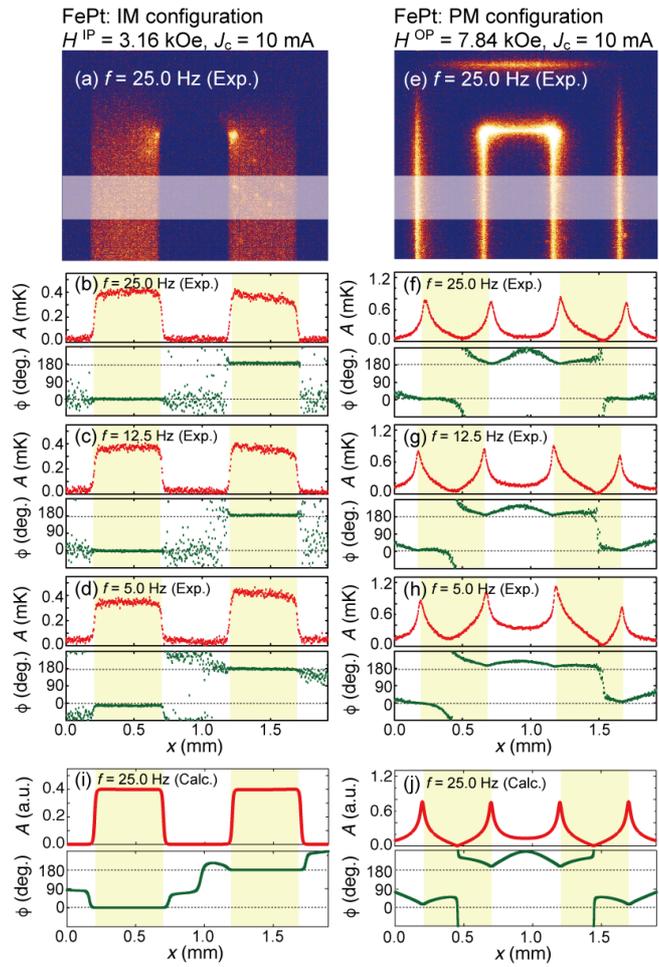

Figure 3

# Visualization of Anomalous Ettingshausen Effect in a Ferromagnetic Film: Direct Evidence of Different Symmetry from Spin Peltier Effect

T. Seki, R. Iguchi, K. Takanashi, and K. Uchida

### -Supplemental Materials-

This Supplemental Materials consist of 2 notes (Supplementary Notes 1-2) and 9 figures (Supplementary Figures S1 - S9).

**Supplementary Note 1: Structural characterization of FePt thin films**

For the present device studying on the anomalous Ettingshausen effect (AEE), the FePt layer was grown on the SrTiO$_3$ (100) substrate at 350ºC. **Figures S1(a) and S1(b)** show reflection high energy electron diffraction (RHEED) images of the SrTiO$_3$ (100) substrate and the FePt layer deposited at 350ºC, respectively. From these RHEED images, it is confirmed that the FePt (100) layer with the flat surface was epitaxially grown on the SrTiO$_3$ (100) substrate. **Figure S1(c)** shows the x-ray diffraction (XRD) profiles of FePt layers. In addition to the profile of FePt deposited at 350ºC, the result for the FePt deposited at 500ºC is shown for comparison. Since the XRD peak angles of FePt are close to those of SrTiO$_3$ substrate, it is difficult for the FePt deposited at 350ºC to distinguish the (001) and (003) superlattice peaks of FePt from the peaks of SrTiO$_3$ substrate. For the FePt deposited at 500ºC, on the other hand, the superlattice peaks appear at the high angle side of the substrate peaks because the high temperature deposition leads to the promotion of $L1_0$ ordering. Thus, the intensities of superlattice peaks are increased for the FePt deposited at 500ºC. Considering that the FePt deposited at 500ºC shows the clear superlattice peaks, the substrate temperature of 350ºC is not enough high to largely promote the $L1_0$ ordering.

In summary, the FePt deposited at 350ºC is the (001)-oriented epitaxial layer, but its degree of long-range order is low, *i.e.* partially $L1_0$-ordered FePt.

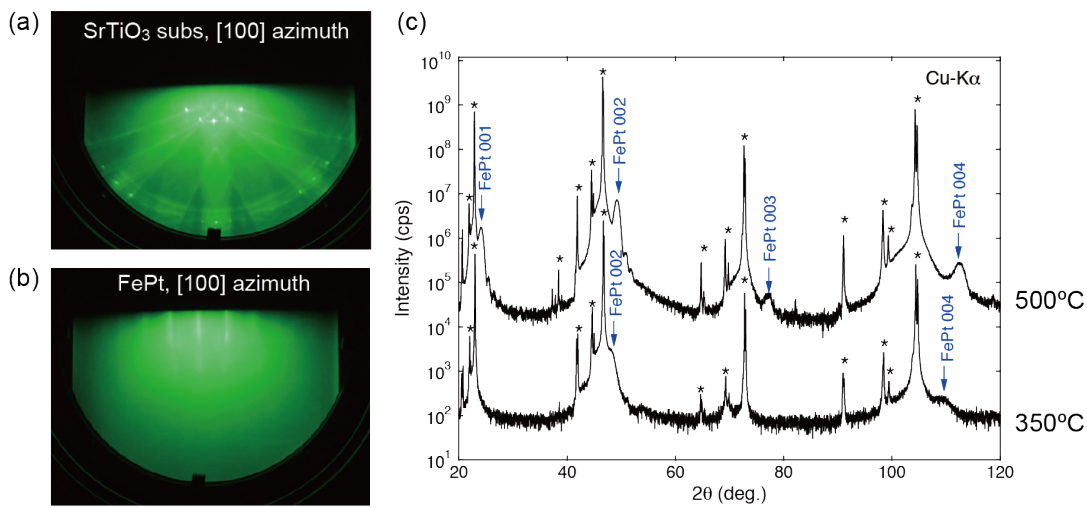

**Supplementary Figure S1:** RHEED images and XRD profiles for the FePt on the SrTiO$_3$ substrate. (a) RHEED images of the SrTiO$_3$ (100) substrate and (b) the FePt layer deposited at 350ºC. (c) XRD profiles with Cu-$K\alpha$ radiation for FePt layers deposited at 350ºC and 500ºC. The asterisks denote the reflections from the SrTiO$_3$ (100) substrates.

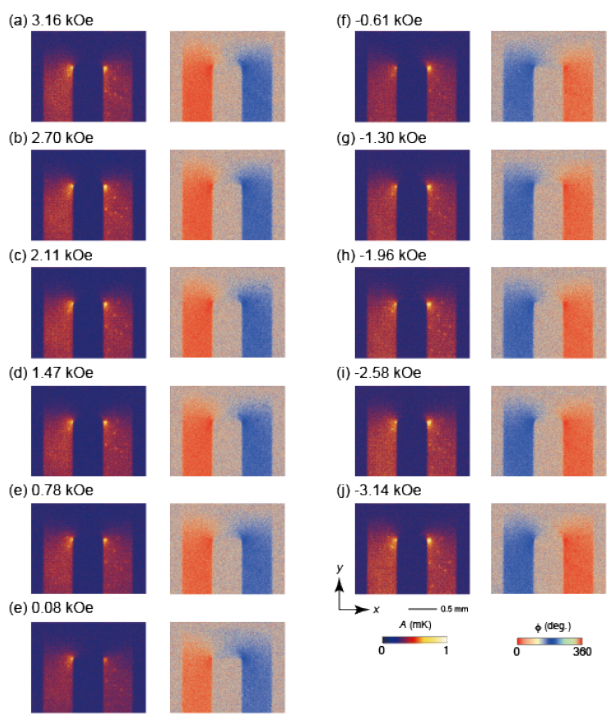

**Supplementary Figure S2:** Thermal images for the FePt as a function of in-plane magnetic field ($H^{IP}$). $H^{IP}$ was swept from 3.16 kOe to -3.14 kOe.

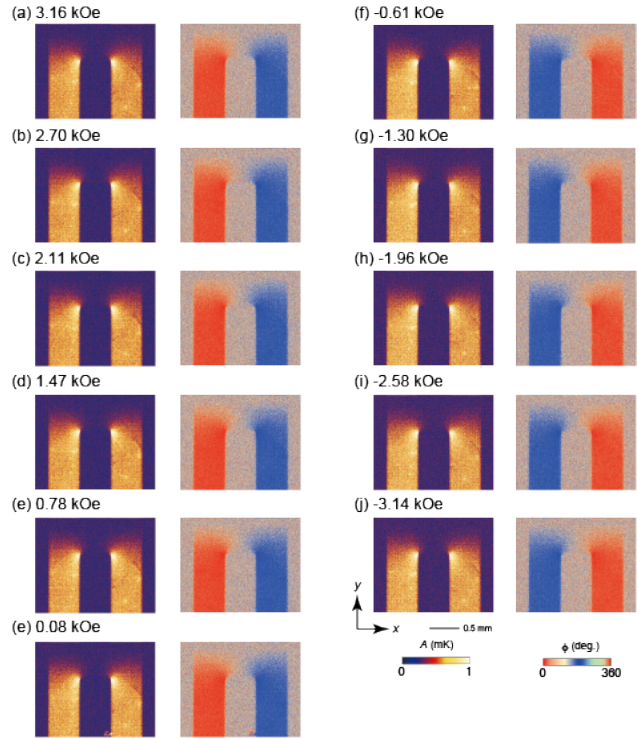

**Supplementary Figure S3:** Thermal images for the Pt / YIG as a function of $H^{IP}$. $H^{IP}$ was swept from 3.16 kOe to -3.14 kOe.

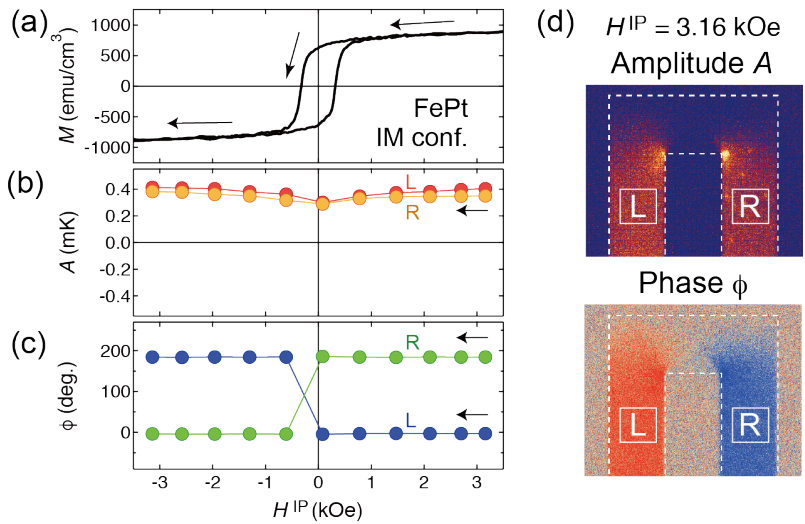

**Supplementary Figure S4:** $H^{IP}$ dependence of (a) magnetization ($M$) for the FePt. (b) The amplitude ($A$) and (c) phase ($\phi$) signals of thermal images on the sample as a function of $H^{IP}$ for the FePt with the IM configuration, where $H^{IP}$ was swept from 3.16 kOe to -3.14 kOe. The data of "L" and "R" are the average values obtained from the areas "L" and "R" enclosed by the white solid lines in (d).

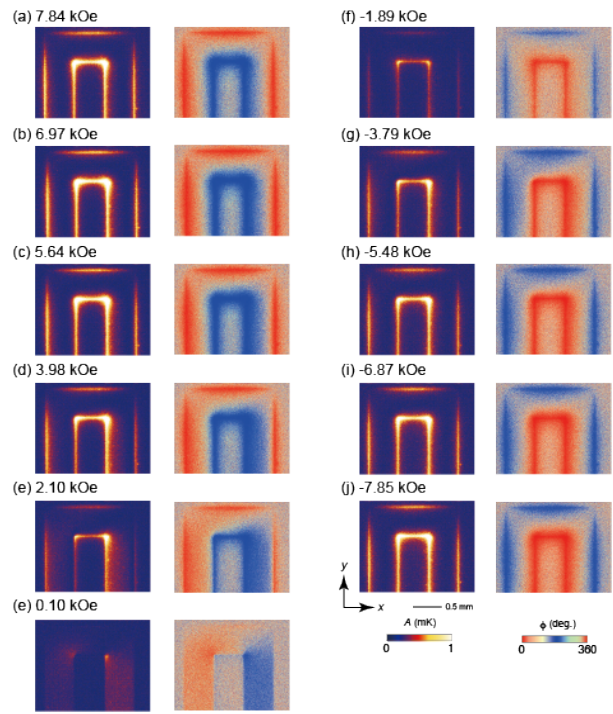

**Supplementary Figure S5:** Thermal images for the FePt as a function of out-of-plane magnetic field ($H^{OP}$). $H^{OP}$ was swept from 7.84 kOe to -7.85 kOe.

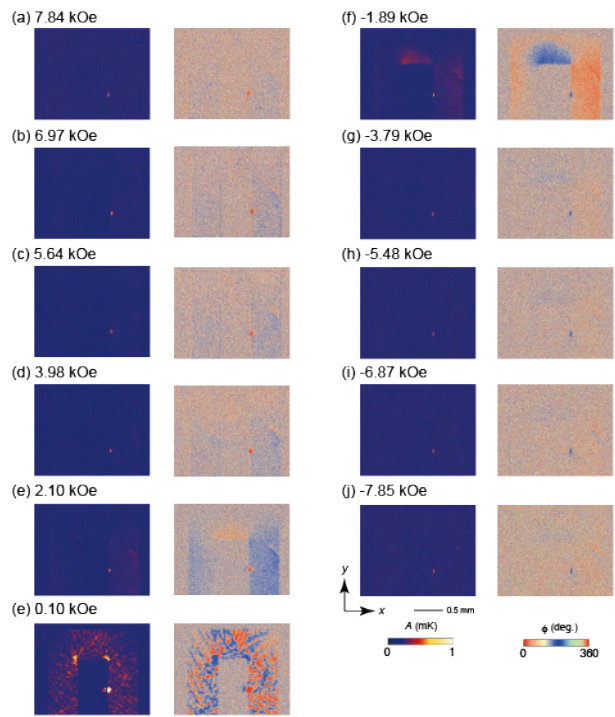

**Supplementary Figure S6:** Thermal images for the Pt / YIG as a function of $H^{OP}$. $H^{OP}$ was swept from 7.84 kOe to -7.85 kOe.

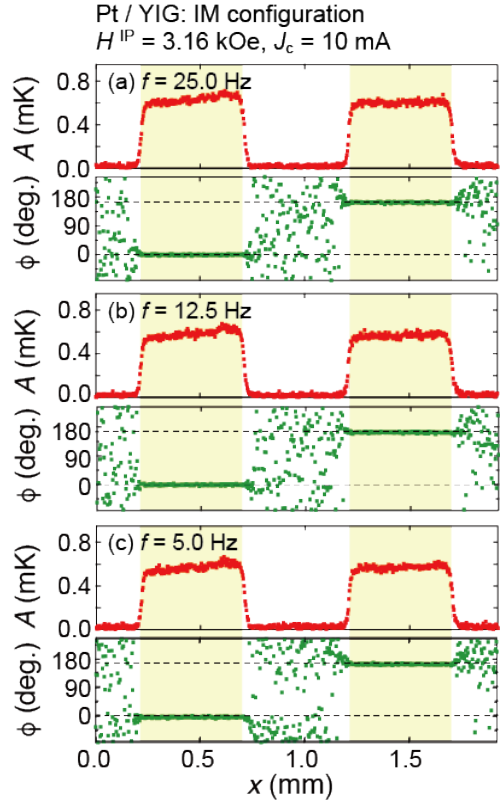

**Supplementary Figure S7:** $x$-directional profiles of $A$ and $\phi$ for the spin Peltier effect of Pt / YIG with the IM configuration ((a) $f = 25.0$ Hz, (b) 12.5 Hz, and (c) 5.0 Hz). $H^{IP}$ was set at 3.16 kOe.

**Supplementary Note 2: Details of numerical modeling based on finite element method**

In order to understand the temperature profiles induced by AEE with the IM and PM configurations, we modeled the distribution of the heat source for AEE using the finite element method [Ref. S1]. We used a two-dimensional model possessing two FePt wires with the SrTiO$_3$ substrate and the black ink. The schematic illustration of cross-sectional view of the model is shown in **Fig. S8(a)**. The thickness and the width of the FePt strip were set at 10 nm and 500 μm, and the gap between the two strips was set at 500 μm. The thicknesses of SrTiO$_3$ substrate and black ink are 500 μm and 10 μm.

The numerical calculation was carried out based on the thermal diffusion equation:

$$C_v \frac{\partial T}{\partial t} = \nabla(k \nabla T) + \dot{Q}, \qquad (S1)$$

where $C_v$ is the volume heat capacity, $k$ is the thermal conductivity, and $\dot{Q}$ is the generated heat. In the case of AEE, the heat current density induced by AEE ($j_q^{AEE}$) is given by

$$\vec{j}_q^{\text{AEE}} = \Pi^{\text{AEE}} \left( \vec{j}_c \times \vec{M} \right), \qquad (S2)$$

where $\Pi^{\text{AEE}}$ is the AEE coefficient. Then, the $\dot{Q}$ is expressed as

$$\dot{Q} = -\text{div}\left(\vec{j}_q^{\text{AEE}}\right). \qquad (S3)$$

In the finite element method calculation, we transformed Eq. (S1) to

$$2i\pi f C_v A \exp(-i\phi) = \nabla\left(k \nabla A \exp(-i\phi)\right) + \dot{Q}_f, \qquad (S4)$$

by substituting $T$ by $\int A \exp\{i(2\pi ft - \phi)\} df$ and $\dot{Q}$ by $\int \dot{Q}_f \exp\{i(2\pi ft)\} df$ in accordance with the alternating input current. Since $\dot{Q}$ appears only at the wire edges, we replaced this term by $+\dot{Q}^{\text{AEE}}$ and $-\dot{Q}^{\text{AEE}}$ localized at the edges of the FePt wires as illustrated in **Figs. S8(b) and S8(c)** for the IM and PM configurations, respectively. The length of minor axis for the heat source is set to be 0.1% (1%) of the wire width (thickness) for the IM (PM) configuration for suppressing the contribution due to the finite length of the heat source. The polarities of the sources for the two wires are set to be opposite. The following parameters were used: $C_v^{\text{FePt}} = 4.5 \times 10^6$ J m$^{-3}$K$^{-1}$ [Ref. S2] and $k^{\text{FePt}} = 17.5$ W m$^{-1}$K$^{-1}$. From the datasheet of SrTiO$_3$ substrate, $C_v^{\text{SrTiO3}} = 2.8 \times 10^6$ J m$^{-3}$K$^{-1}$ and $k^{\text{SrTiO3}} = 11$ W m$^{-1}$K$^{-1}$ were used. For the black ink layer, we assumed $C_v^{\text{Black ink}} = 1 \times 10^6$ J m$^{-3}$K$^{-1}$ and $k^{\text{Black ink}} = 0.5$ W m$^{-1}$K$^{-1}$ considering the value of ZrO$_2$, which is one of the ingredients of the black ink, and the reduction due to porosity. We fixed the temperature for the bottom of SrTiO$_3$ substrate at 300 K, and took into account the heat radiation to air (10 W m$^{-2}$ K$^{-1}$) [Ref. S3].

      **Figures S8(d) and S8(e)** show the temperature modulation in phase with the input charge current in the cross-section with the IM and PM configurations, respectively. Here, the absolute total amount of individual heat sources is set to be the same for both the IM and PM configurations. One can see that from **Fig. S8(d)** for the IM configuration, the temperature modulation in the FePt wires and the above black ink are uniform, and the sharp transition is observed. In contrast to the IM, the PM configuration (**Fig. S8(e)**) shows the remarkable heat loss into the substrate. That is the reason why we obtained the nonlinear temperature profiles for the PM configuration. The calculated $\Delta T^{\text{PM}} / \Delta T^{\text{IM}}$ is ~ 4 at $f$ = 25.0 Hz, where the values are estimated from those at the surfaces of black ink layers. As this factor mainly depends on the thermal property of the substrates, the factor for the Pt / YIG system is nearly the same considering the similar thermal properties of YIG to those of SrTiO$_3$ [Ref. S4], meaning that the observed absence of $\Delta T^{\text{PM}}$ in the Pt / YIG is indeed due to the symmetry of the SPE.

      We also confirmed the steady-state temperature modulation due to the AEE and its

deviation when $f$ changes. **Figure S9** summarizes $f$ dependences of calculated $|\Delta T|$ and $\phi$ signals for the IM (**(a) and (b)**) and the PM configurations (**(c) and (d)**). The data were calculated at the surfaces of FePt wire and black ink. The experimental $f$ dependences of $|\Delta T|$ for the IM and PM configurations are also shown in **Figs. S9(e) and S9(f)**, respectively, for comparison. The IM configuration (**Figs. S9(a) and S9(b)**) shows no remarkable $f$ dependence of $|\Delta T|$ although the delay of $\phi$ is observed as $f$ is increased because of the influence of thermal diffusion. In the case of PM configuration (**Figs. S9(c) and S9(d)**), on the other hand, $|\Delta T|$ gradually decreases with $f$ for both the surfaces of FePt wire and black ink. This is also related to the thermal diffusion. Importantly, the calculated $f$ dependences of $|\Delta T|$ signals are qualitatively consistent with the experimental results.

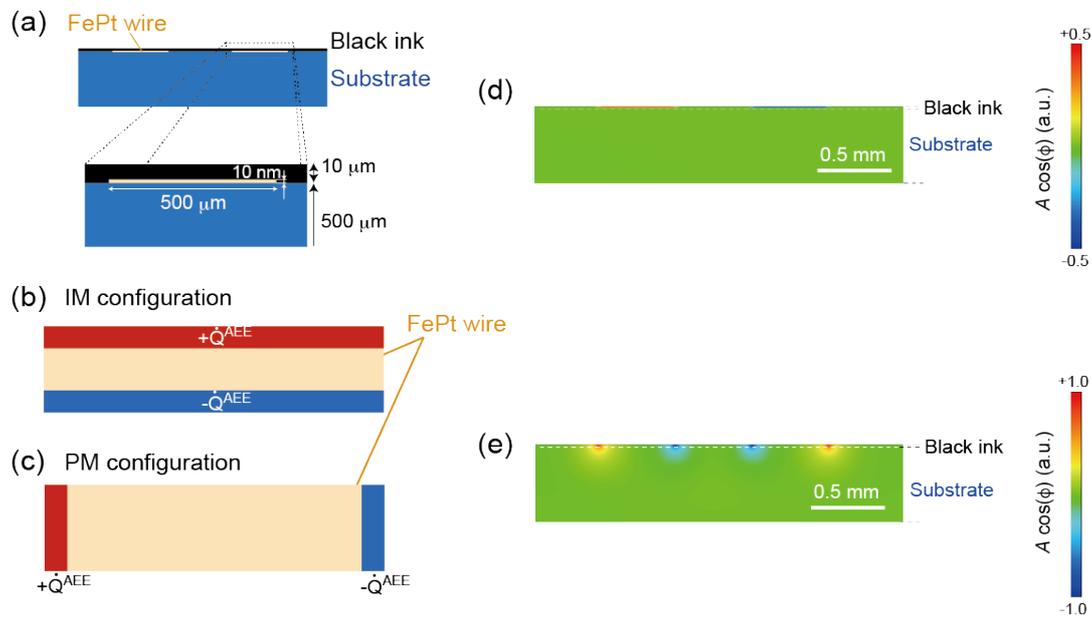

**Supplementary Figure S8:** (a) Schematic illustrations of cross-sectional view of the calculation model using the finite element method. Illustrations for the locations of heat sources ($+\dot{Q}^{AEE}$ and $-\dot{Q}^{AEE}$) for (b) IM and (c) PM configurations. Real part of complex temperature modulation ($A\cos(\phi)$) in the cross-section with the (d) IM and (e) PM configurations, where $f$ was set at 25 Hz.

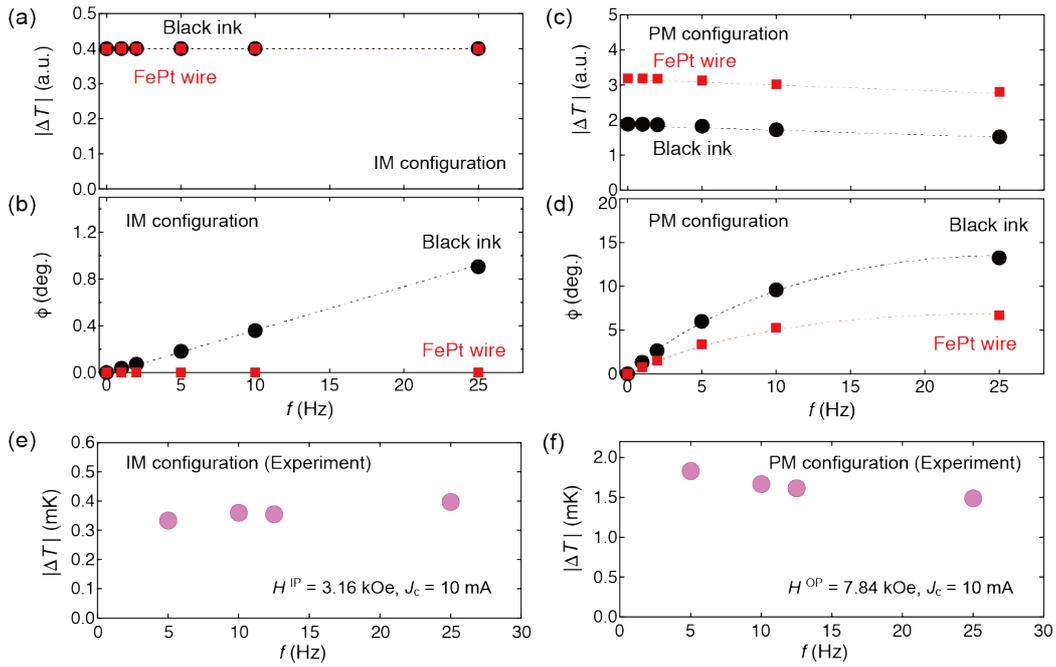

**Supplementary Figure S9:** (a) $f$ dependences of calculated $|\Delta T|$ and (b) $\phi$ signals for the IM configuration. (c) $f$ dependences of calculated $|\Delta T|$ and (d) $\phi$ signals for the PM configuration, where $|\Delta T|$ was estimated from the temperature modulation difference between the inside edge and the outside edge of the left FePt wire. The solid square and the solid circles denote the data obtained at the surfaces of FePt wire and black ink, respectively. (e) Experimental $f$ dependence of $|\Delta T|$ for the IM configuration. (f) Experimental $f$ dependence of $|\Delta T|$ for the PM configuration, where $|\Delta T|$ was obtained from the temperature modulation difference between the inside edge and the outside edge of the left FePt wire.